\documentclass[printer]{aa} 

\usepackage{graphicx}
\usepackage{txfonts}
\begin{document}

   \title{Massive young stellar objects in the N66/NGC346 region of the SMC {\thanks {Based on observations collected at the European Organisation for Astronomical Research in the Southern Hemisphere 
under ESO programme 063.C-0329(A) and 072.C-0466(A).}}}

   \author{M. Rubio
          \inst{1}
          \and
         R H. Barb\'a \inst{2}
          \and
        V. M. Kalari \inst{1}
          } 

   \institute{Departamento de Astronom\'{\i}a, Universidad de Chile,
       Casilla 36-D Santiago, Chile\\
              \email{mrubio@das.uchile.cl}
         \and
             Departamento de F\'{\i}sica, Universidad de La Serena, Benavente 980,
       La Serena, Chile\\
             }

   \date{Received September 15, 1996; accepted March 16, 1997}

 
  \abstract
{We present {\it HK} spectra of three sources located in the N66 region of the Small Magellanic Cloud. The sources display prominent stellar Br$\gamma$ and extended H$_{2}$ emission, and exhibit infrared excesses at $\lambda$\,$>$\,2$\mu$m. Based on their spectral features, and photometric spectral energy distributions, we suggest that these sources are massive young stellar objects (mYSOs). The findings are interpreted as evidence of on-going high mass star formation in N66.}

   \keywords{Magellanic Clouds --- 
HII regions --- 
infrared: stars ---
techniques: spectroscopic ---
ISM: individual (N66) ---
stars : pre-main-sequence
               }

   \maketitle
%

\section{Introduction}

The metal-poor Magellanic Clouds with their unobscured line of sight, their small
internal extinction and their profusion of star-forming regions, constitute
one of the best laboratories to study the interaction between forming massive stars
and their environment.
In particular, the Small Magellanic Cloud (SMC), having a metallicity ($Z$) $1/5$ solar
metallicity (Russell \& Dopita 1992), allows one to study the process of star formation in a gas-rich 
environment approaching that of the early Universe.  Its proximity, at a distance of 61$\pm$1\,kpc (Hilditch et al. 2005), makes it possible 
to resolve individual stars in a low metallicty environment, and study star formation in metal-poor environments (Kalari \& Vink 2015; Kalari et al. 2018).

 

N66 is the brightest H\,{\sc ii} region in the SMC \citep{Henize1956} and is located
in the northeast of the SMC bar. It is excited by the dense cluster of massive young
stars, NGC\,346. 
\cite{Massey1989} performed an extensive study of the stellar content of this
cluster, also referred as N66A, and identified at least 33 O-type stars,
including 11 of spectral type O6.5 or earlier. 
Twenty-two of these O stars are contained in NGC\,346 and the others are
isolated or in small groups. 
Multi-wavelength studies of N66 have shown the complexity and richness of
this region. 
\cite{Contursi2000} showed mid-infrared (mid-IR) emission peaks coincident
with the main features of the ionized gas. 
\cite{Rubio2000} detected weak molecular gas towards the H\,{\sc ii} 
region from high-sensitivity CO observations, and associated to compact dense molecular
hydrogen knots from emission in the v(1-0) S(1) line of H$_2$ at 2.12\,$\mu$m. 
These data revealed a beautiful photodissociation region with molecular gas
not yet photo-dissociated by the UV radiation of the stars, concentrated in
dense clumps with a very small surface filling factor. 
In addition, \cite{Rubio2000} discovered several embedded stars and/or star
clusters, and suggested that three successive stellar generations have already formed in less than 3 million years. 
 
Studies of the stellar population done with \textit{HST/ACS} observations by
\cite{Nota2006}, Gouliermis et al. (2006), \cite{Sabbi2007}, Schmeja et al. (2009) have found a large number of
pre-main-sequence stars with ages between 3-5 Myrs in N66. Dedicated {\it Spitzer} studies have uncovered massive young stellar objects (mYSOs) in the region (Simon et al. 2007). The spatial distribution of optical and infrared stellar populations shows a concentration of stars coincident with 
the neutral molecular clumps identified by \cite{Rubio2000} from CO(2-1)
observations. 
The multi-wavelength photometric studies strengthened the suggestion that the region may have sites of
recent or ongoing star formation, and that the newly formed stars are still
embedded in the molecular cloud and thus invisible at optical wavelengths. 

In this paper we present, {\em H} and {\em K} band
spectroscopy of three early type stars in N\,66 in the SMC. These three stars represent the best candidate mYSOs in N\,66  based on their near-infrared {\it JHK}s colours. They are bright and have the largest infrared (IR) excess of the sample. The near-infrared spectral observations complement the multi-wavelength photometric studies of YSOs (young stellar objects) conducted in this region, and help gain a more detailed understanding of their nature, and differences with respect to Galactic counterparts.

This paper is organised as follows. Section 2 contains a description of the photometric and spectroscopic observations of our mYSOs. Section 3 present our results from these observations. The discussion pertaining to the nature, and formation scenario of the mYSOs is given in Section 4. Finally, we present our conclusions in Section 5.

\section{Observations}

\subsection{Near-infrared photometry and target selection}

We obtained deep (near-infrared) nIR {\it JHK}$_{\rm s}$ images of N66 using the Infrared Spectrometer and array Camera (ISAAC) imager mounted on the Antu 8.2m Very Large Telescope (VLT) at Paranal Observatory  We used the short wavelength arm equipped with a $1024\times 1024$ pixels
Hawaii Rockwell array, with a pixel scale of $0\farcs148$ pix$^{-1}$
and a total field of view (FOV) of about $2\farcm5\times2\farcm5$. 
The observations were done in a series of 6 frames, each individual
frame with 10s  integration time in each filter. The median seeing of our observations was 0.8$\arcsec$.
The individual frames were coadded and a 60s  image was stored. 
We observed two fields with an overlap of about 30\arcsec\ to cover most
of the N66 region. 
In each filter, every 10 minutes of on-source imaging we interleaved sky frame
observations. 
These sky frames were chosen in a field with faint stars and no extended
emission at 300\arcsec\ South of the position of N66. 
The sky field was observed in a similar way as the source frames.
The procedure was repeated several times to achieve a final integration time
of 3600s  in $J$, $H$ and $K_s$.

Previous to the VLT observations, we had conducted nIR $JHK_s$
imaging of N66 using the 2.5-m Du Pont Telescope at Las Campanas
Observatory with the $256\times256$ NICMOS III
Camera, ClassicCAM \citep{Persson1992}. 
These images covered a $1\arcmin\times1\farcm2$ area with a pixel scale of
$0\farcs35$ pix$^{-1}$ and typical seeing of $0\farcs8$ in $K_s$.   
The observation consisted of a series of 10 frames, each individual 
frame with an integration time of 20s  per filter. 
Sky frames in each filter were taken in a field at $300\arcsec$ South with
faint stars and no extended emission.  
The sky field was observed in a similar way as the source frames.
This observing procedure was repeated to produce a 9 position mosaic 
with separation of $20\arcsec$. 
The total integration time in each filter resulted in 1800s  in
$K_s$, 1800s  in $H$, and 2000s  in $J$.

\begin{figure*}
\centering
\includegraphics[width=15cm, angle=0]{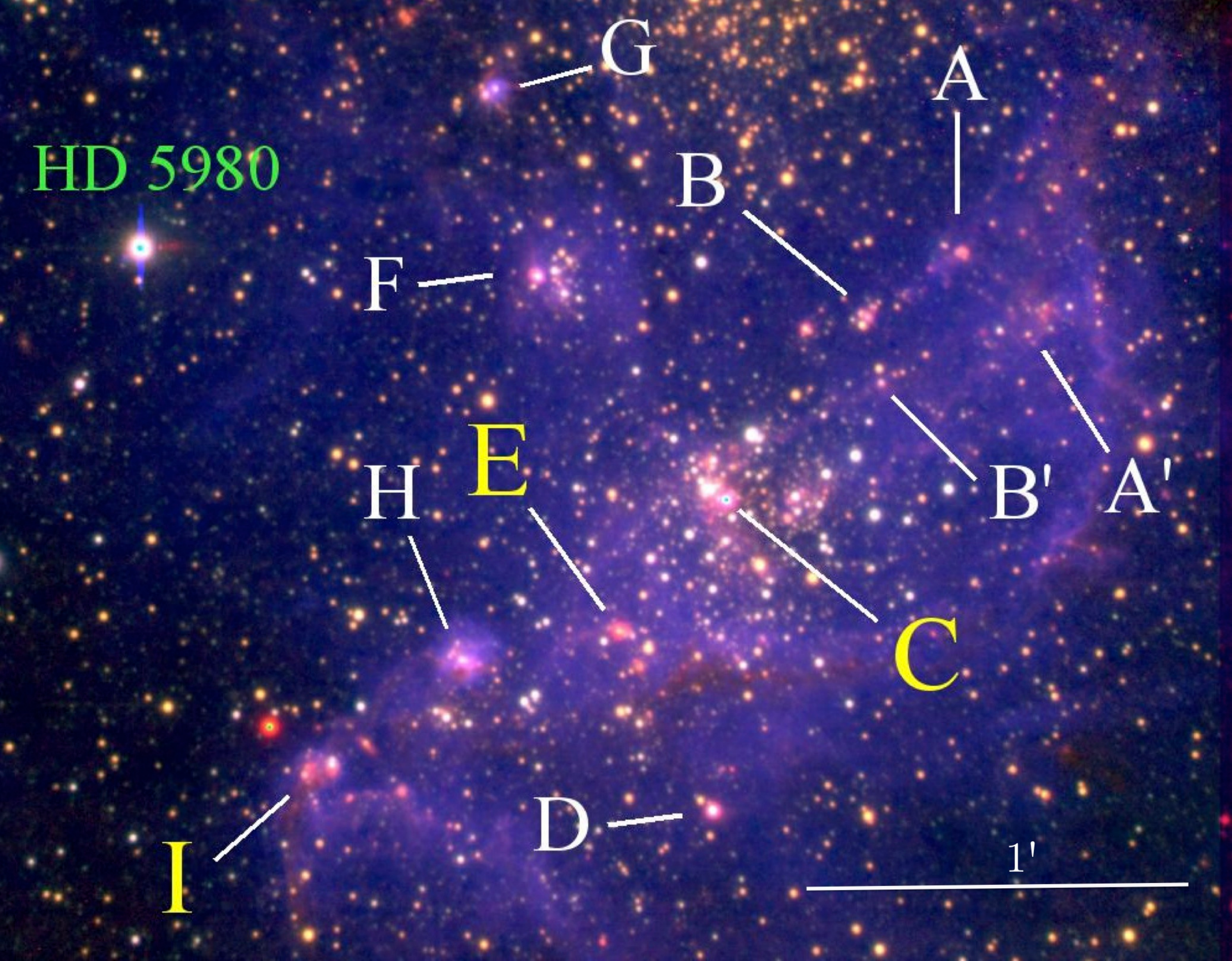}
\vskip0.1cm

\caption{ Colour composite image of N66. 
In red and green channels are $K_s$- and $J$-band images obtained with
VLT/ISAAC, respectively. Blue channel is an VLT/FORS1 archival H
$\alpha$
image. 
Indicated in the images are the ISO sources labelled A to H \citep{Rubio2000};
\citep{Contursi2000}.  
Yellow labels C, E, I indicate the nIR sources of this study. North is up, and east is to the left. Scalebar on the lower left gives 1 arcmin. 
} 
\label{fig1}
\end{figure*}

To produce the final images, for both LCO and VLT observations, each image
was dark corrected, flat-fielded and sky subtracted, and then median combined
using IRAF {\footnote{IRAF is distributed by the National Optical Astronomy Observatory, which is operated by the Association of Universities for Research in Astronomy (AURA) under a cooperative agreement with the National Science Foundation.
}} procedures.
The final images were aligned with respect to the $K_s$ image by means of
several common stars.  
The final VLT image, which combines the two overlapping fields cover a total
$2\farcm5\times5\farcm0$ area.   
The resulting LCO image, of only $65\arcsec\times76\arcsec$ size includes the
NGC\,346 cluster and the ISO peaks E and I \citep{Contursi2000}. Figure~\ref{fig1} shows a composite H$\alpha$, $J$, $K_s$ image of the central
$3\farcm4\times2\farcm7$ area of N66 produced with the $K_s$ and $J$ images
obtained at VLT and a FORS/VLT H$\alpha$ image retrieved from the ESO archive,
as red, green, and blue channels, respectively. Several IR sources appear very bright in this composite image. 
These sources are located towards the H$_2$ knots and 7\,$\mu$m peaks
associated with CO(2-1) emission from molecular clouds (\cite{Rubio2000},
\cite{Contursi2000}).

We performed point-spread-function photometry on ISAAC/VLT images using
DAOPHOT(Stetson 1989) incorporated in IRAF, reaching a limiting
magnitude of about $K_s=22$, $H=21$ and $J=21$, respectively. 
The photometric calibration was made with respect to 2MASS photometry. We choose the 2MASS system since it resembles the ISAAC and LCO near-infrared filters. We identified 20 bright stars outside the central cluster which were common in the LCO, ISAAC and 2MASS images. We then applied offsets determined with respect to the 2MASS photometry to the LCO and ISAAC photometry. 
The brightest IR sources
appear saturated in the VLT images so  we  performed  our own photometry and  used the LCO $JHK_s$ images  to measure the IR 
magnitudes of these brightest sources  instead of using  Gouliermis et al. (2010) photometry.  Table~\ref{mags} gives the $JHK_s$ magnitudes obtained for the three selected objects. In Figure ~\ref{cmd}, in the top panel, we show a colour--magnitude diagram of
the stars of the region shown in Fig.~1 and brighter than $J=20$, obtained
from photometry performed on the VLT images.  
In the same figure we show in the lower panel the ($J-H$) vs. ($H-K$s) colour--colour diagram. 

\begin{table*}
\begin{center}
\caption{The nIR magnitudes of mYSOs in N66.\label{mags}}
\begin{tabular}{crrrrrrrrr}
\hline\hline
Source & RA(2000) & DEC(2000) & J & ${e_J}$ & $H$ & $e_H$ & $Ks$ & $e_{Ks}$ &
Rubio et al. (2000)\\
\hline
1 & $00\,59\,05.43$ & $-72\,10\,35.5$ & 14.44 & 0.02 & 13.50 & 0.04 & 12.05
(L) & 0.05 & C\\
2 & $00\,59\,09.29$ & $-72\,10\,57.4$ & 16.37 & 0.03 & 14.66 & 0.02 & 12.86
(L) & 0.01 & E\\
3 & $00\,59\,19.66$ & $-72\,11\,19.6$ & 16.01 & 0.03 & 15.61 & 0.04 & 15.15
& 0.04 & I\\
\hline
\end{tabular}
\tablefoot{The (L) denotes those magnitudes obtained from photometry
  performed on the LCO/du Pont images, these sources appear saturated in the
  ISAAC/VLT $K_s$ images.} 
\end{center}
\end{table*}

\subsection{Infrared spectroscopy}

\begin{figure}
\centering
\includegraphics[width=8cm, angle=0]{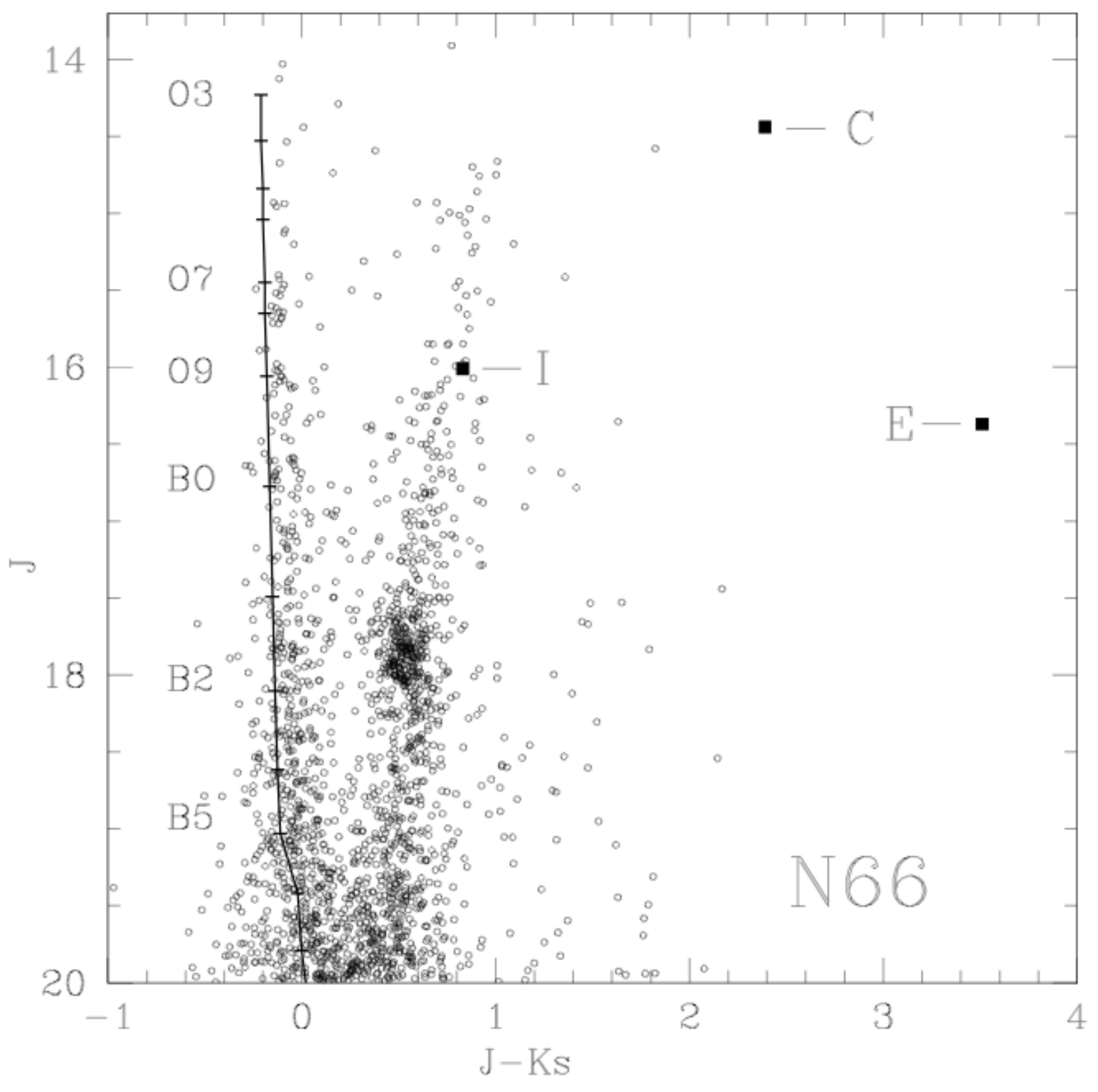}
\includegraphics[width=8cm, angle=0]{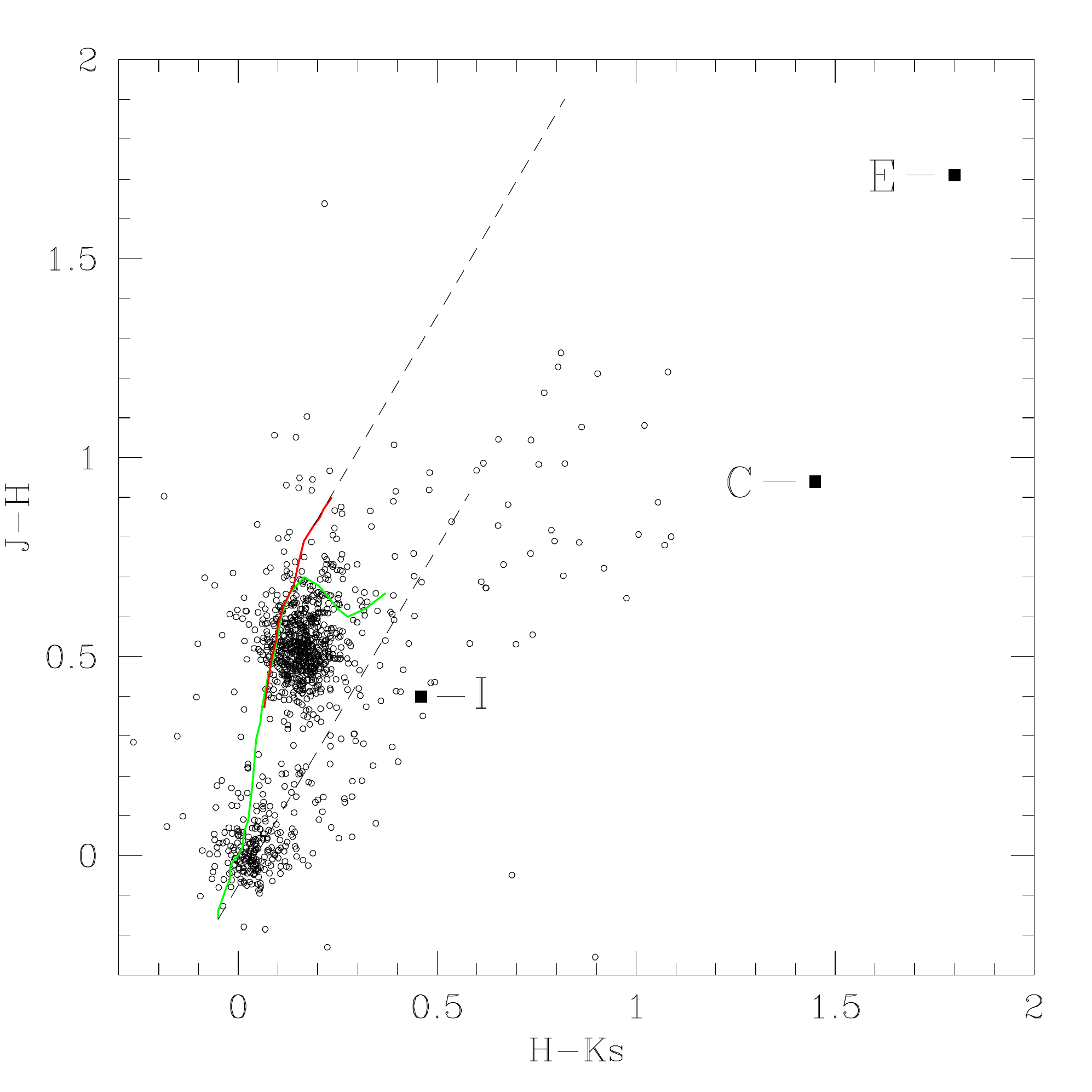}
\caption{VLT/ISAAC nIR colour--magnitude and colour--colour diagrams of N66, with the positions of the selected sources for spectroscopy marked. For source C and E, the $K$s magnitudes are from LCO observations. The three sources correspond to the the brightest infrared excess sources in their respective molecular clouds. The green and red solid lines in the colour--colour diagram are the dwarf and giant locus respectively. The reddening track of a normal O3 V star and Cool Giant G is plotted as  dash black lines. The ZAMS between B5 and O3 at the distance of the SMC is indicated with a solid black line.}
\label{cmd}
\end{figure}

We selected the three brightest sources for $K$-band spectroscopy  based on the nIR excess from the images obtained with ClassicCAM
attached to the Du Pont telescope at Las Campanas Observatory and with  ISAAC/VLT at Paranal Observatory.
 Fig.~\ref{cmd} show these sources which  correspond spatially to the emission peaks designated C, E and I in \cite{Contursi2000}.

We obtained low-dispersion ($R\sim1000$), {\em HK}-band spectra of these three
sources in N66 in the SMC using the SofI instrument at the NTT at La Silla
Observatory. The total integration time  on each source was 1200s and the signal-to-noise
ratio ($S/N$) obtained ranged from 30 to 80.  
Sky selection for background subtraction was crucial as all the selected
sources were embedded in bright nebulosity and crowded fields.  
The observations were conducted in an ``ABBA'' sequence.
Darks and flat-field calibrations were obtained for each night along with
several spectrophotometric standards to get flux calibrations. Telluric correction was performed using the standards Hip4536 or Hip7806 (F5V and F7V spectral types respectively), taken immediately before and after the target, at similar airmass. The velocity is given in heliocentric rest frame.
The nIR spectra of the sources are shown in Figure~3. 
The spectra of sources C and E were obtained along the slit at the same time with a slit position-angle of $PA=39^\circ$.
The spectra of source I was obtained with the slit in $PA=59^\circ$.

\subsection{Archival photometry}

In addition to the infrared photometry campaign, we used the archival {\em Hubble} imaging study of Sabbi et al. (2007) to find optical counterparts in the filters F555W (hereafter $V$) and F814W (hereafter $I$). For the stars C and I, we find cross matches within a radius of 0.1$\arcsec$, acceptable for the crowding and confusion in this region. In the 0.06$\arcsec$ HST resolution images there are two stars  associated to our Source C. These stars are visible in $V$ and their {\it HST} $VI$ photometry show  two roughly similar brightness stars in $VI$ within 0.5$\arcsec$ of the source C, catalogued by Sabbi et al. (2007) with  ids  152 and 168 with V magnitude  of 16.62, 16.71, respectively. We select ids 152 as our Source C counterpart, since it falls within the cross match radius (0.5$\times$ the seeing of the infrared imaging) of the Source C infrared coordinates. The infrared photometry of Source C includes a contribution from both HST sources. For Source E, the nearest optical neighbour is $>$0.15$\arcsec$ and does not seem to be a physical counterpart of the nIR source. 


We used the HST  H$\alpha$ image to check for emission line stars and compact HII regions and we cross match the stars with mid-infrared {\em Spitzer} (Simon et al.  2007) and {\em Herschel} public catalogues (Seale et al. 2014). We have checked that their are no neighbours falling within the FWHM of the imaging at mid-IR wavelengths, thereby contaminating the determined magnitudes. The photometry of the {\em Spitzer} IRAC 3.6$\mu$m--8$\mu$m bands are given in Table \ref{mags2}, while the fluxes of the long wavelength {\em Spitzer} MIPS 24-70$\mu$m, and {\em Herschel} 100-350$\mu$m photometry is given in Table \ref{mags3}.

\begin{table*}
\begin{center}
\caption{Archival optical and near-infrared photometry\label{mags2}}
\begin{tabular}{crrr}
\hline\hline
Filter & Source C & Source E &  Source I\\
        & (mag) & (mag) & (mag)\\
\hline
F555W$^1$ & 16.617\,$\pm$0.009 & -- & 16.919\,$\pm$0.011\\
F814W$^1$ & 15.983\,$\pm$0.008 & -- & 16.865\,$\pm$0.012\\
 & (0.065$\arcsec$)$^*$ & -- & (0.054$\arcsec$)$^*$ \\
3.6$\mu$m$^2$ &   10.101\,$\pm$0.018 & 10.063\,$\pm$0.014 & 12.842\,$\pm$0.179\\
4.5$\mu$m$^2$ &    9.422\,$\pm$0.013 & 8.861\,$\pm$0.016  & 12.142\,$\pm$0.034\\
5.8$\mu$m$^2$ &    8.714\,$\pm$0.015 & 7.824\,$\pm$0.009  & 10.391\,$\pm$0.035\\
8.0$\mu$m$^2$ &   7.234\,$\pm$0.012  & 6.811\,$\pm$0.008  & 8.867\,$\pm$0.047 \\
 & (0.039$\arcsec$)$^*$ & (0.067$\arcsec$)$^*$ & (0.416$\arcsec$)$^*$ \\
\hline
\end{tabular}
\tablefoot{$^1$ Sabbi et al. 2007; $^2$ Seale et al. 2014; $^*$Denotes the separation in arcsec of our source postions} 
\end{center}
\end{table*}

\begin{table*}
\begin{center}
\caption{Archival mid-infrared photometric fluxes \label{mags3}}
\begin{tabular}{crrr}
\hline\hline
Filter$^1$ & Source C & Source E &  Source I\\
        & (mJy) & (mJy) & (mJy)\\
\hline
24$\mu$m &   2154\,$\pm$8.48 & 944.5\,$\pm$4.13 & 425.5\,$\pm$4.3\\
70$\mu$m &    3766\,$\pm$51.51 & --  & 5598\,$\pm$55.33\\
100$\mu$m &    1695\,$\pm$225.9 & 2400\,$\pm$208.7  & 3083\,$\pm$214.5\\
160$\mu$m &   1724\,$\pm$199.3  & 1461\,$\pm$181.9  & 2702\,$\pm$213.6 \\
250$\mu$m &    822.1\,$\pm$57.74 & 502.8\,$\pm$50.12  & 1118\,$\pm$46.24\\
350$\mu$m &   302.1\,$\pm$48.99  & 442.9\,$\pm$53.73  & 448\,$\pm$39.20\\
 & (2.14$\arcsec$)$^*$  & (0.487$\arcsec$)$^*$  & (1.083$\arcsec$)$^*$  \\
\hline
\end{tabular}
\tablefoot{$^1$ Seale et al. 2014; $^*$Denotes the separation in arcsec of our source postions} 
\end{center}
\end{table*}

\section{Results}

\subsection{Photometry}

The three IR sources for which we have obtained spectra are plotted and labelled 
in the $(J-K_s, K_s)$ colour--magnitude diagram in Figure~\ref{cmd}.
In this diagram, the solid line represents the location of the 
zero-age main sequence (ZAMS; Hanson et al. 1997), using a distance 
modulus of 18.9 to the SMC and assuming no reddening (Keller \& Wood 2006).
Source C is one of the most luminous IR sources in NGC\,346 (after WR/LBV
system HD\,5980) with $K_s = 12.05$, and source E displays the reddest colour
with a $J-K_s=3.4$. 
Both sources show IR colours distinct to those of the majority of the stars in
the field and consistent with their different nature.  
Source C and E lie in the region of the colour-colour diagram of IR excess (see 
Figure~\ref{cmd}) and are similar to the brightest IR sources detected in
30~Doradus in the LMC (Rubio et al. 1998). Source C has an absolute Ks magnitude of $M_{Ks}=-6.88$, Source E,
$M_{Ks}=-6.07$, and Source I $M_{Ks}=-3.78$, assuming 
no reddening to determine their absolute magnitudes. 
If they do suffer any extinction then they would become even brighter.
 
According to the distribution of massive YSOs in the  $(J-K_s, K_s)$ 
colour--magnitude diagram (Figure~1 in Bik et al. 2006) sources C and E
are extreme bright objects comparable in brightness to those of the 
Galactic massive YSOs in W31 (Blum et al. 2001) or NGC\,2024 
(Lenorzer et al. 2004), and S106-IRS4 (Felli et al. 1984), and much brighter
than Herbig Be stars (Eiroa et al. 2002; Kalari et al. 2014). 
Thus, they are extreme bright young stellar objects. 
Both sources have magnitudes $ K_s $ similar to those mYSO observed in the LMC 
by Oliveira et al. (2006): N \,157B \,S1, and N \,105 \,S1, with $ K_s $
11.58  and 13.31 magnitude, respectively, or for IRAS \,$ 05328-6827$ with $K_s = 11.98$ (van Loon et al. 2005). 
For the  SMC there is very little nIR spectroscopic information of mYSOs.  L band spectroscopy   
for three mYSO candidates has been reported by van Loon et al. (2008)
one of which, IRAS \,$01039-7305$ has a $K_s = 11.69$ magnitude similar to Source C.

\subsection{Near Infrared spectral features}


We have identified the emission lines in the nIR spectra of each source. These are given in Table 4 for Source C, Table 5 for Source E and Table 6 for source I. In
 the tables we give the wavelength, flux and equivalent width (EW) for each identified line. We describe the spectral features of the three sources in the following:

\begin{figure*}
\label{sed}
\centering
   \includegraphics[width=10cm]{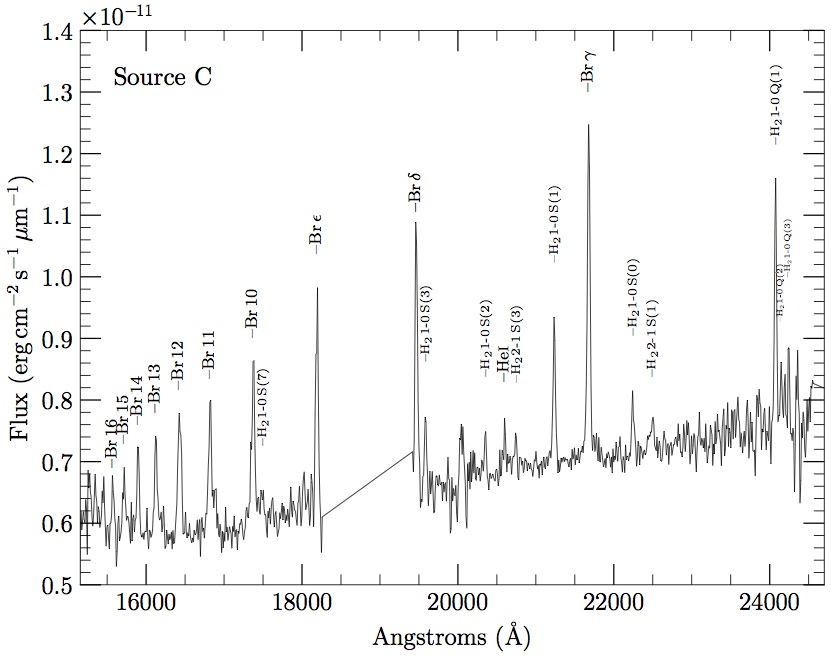}
   \includegraphics[width=10cm]{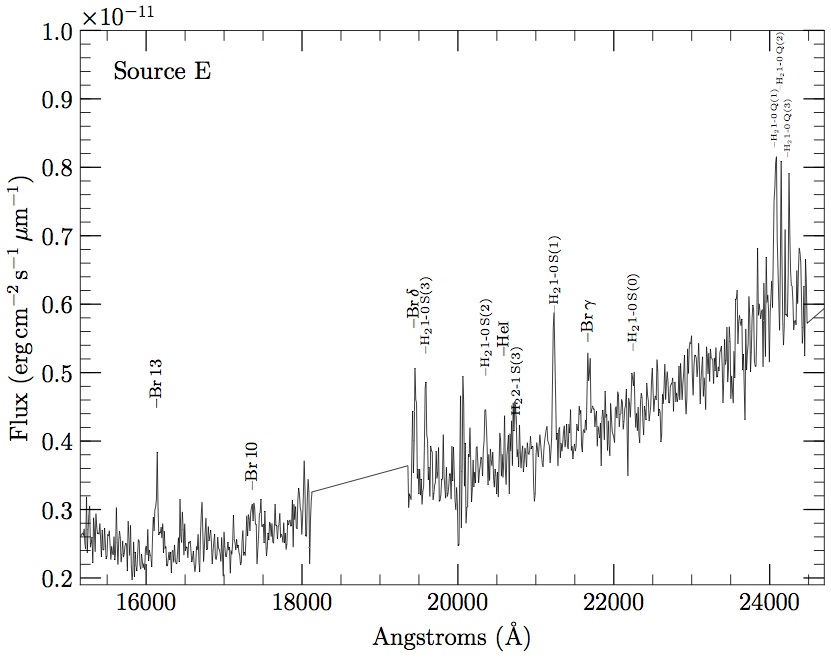}
   \includegraphics[width=10cm]{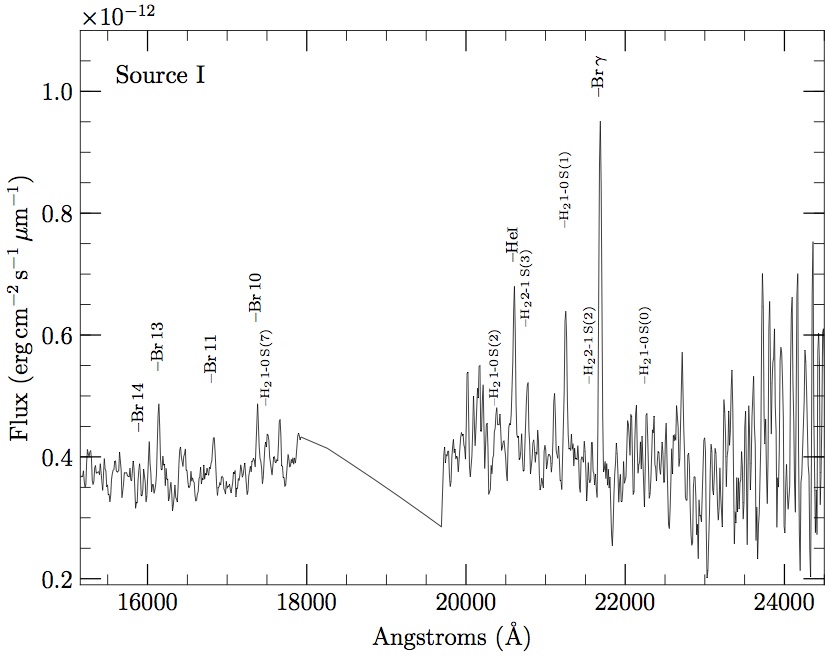}
\caption{{\it HK} spectra of each star as labelled. Prominent H Brackett series, He, and H$_2$ emission lines are seen. }
\label{sed1}
\end{figure*}

\noindent
{\it Source C}: 

The spectra of Source C (Fig. 3) suggests an embedded star within a compact ($\lesssim$\,5$\arcsec$) nebulosity. 
In the 2-D long slit image, we find extended emission centred on the source, being
brighter in H$_2$ molecular emission than the Br$\gamma$ emission. The spectrum shows a rising continuum towards longer wavelengths with H$_2$ emission lines. The spectrum displays prominent HeI emission line at 2.058\,$\mu$m, indicating that the source must have a temperature $>$20\,000 K to excite this line. This emission is not extended (in contrast to Br$\gamma$ or H$_2$), therefore probably comes from the source itself, and suggests the source must be an early type star. According to Trundle et al. (2007) who approximate a spectral type-effective temperature relation for the SMC metallicity,  the star is earlier than at least B3 to have temperatures high enough to excite the HeI 2.058\,$\mu$m line. The absence of any HeII lines along with the CIV 2.070-2.079 \,$\mu$m lines suggests a later O-early B spectral type. But, the emission line spectra of such sources do not allow for a highly reliable classification using purely the MK scheme. Based on the approximate spectral type between late O (O7)-early B (B3) at SMC metallicity, we take the logarithm of the effective temperature (log\,$T_{\rm eff}$) of the source to be 4.5\,$^{+0.1}_{-0.2}$ (Trundle et al. 2007). The error on the temperature is due to the spectral typing error of 6 sub types.

{\it Source E}: is the second brightest ISO peak (Contursi et al. 2000) and shows the most extreme nIR colours (Table~\ref{mags}).In general, the spectrum is comparatively featureless to Source C or I, with  molecular H$_2$ and ionized hydrogen emission lines. The ratio of H$_2$ $2.12\,\mu$m to Br$\gamma$ emission lines is higher than that of source C. Also, the long-slit spectrum shows extended emission in H$_2$ of about $3\arcsec$, and the Br$\gamma$ emission line is stronger on the source but fainter in the nebulosity.   
The log\,$T_{\rm eff}$ assuming a B0 spectral type at SMC metallicity is 4.45\,$^{+0.1}_{-0.2}$ \citep{Massey2005}.  

{\it Source I}: is located in an H$_2$ knot (Rubio et al. 2000, Figures 6 and 8), embedded in the head of a trunk-like feature pointing to the direction of the core of NGC\,346. The nIR spectrum  shows the H$_2$, Hydrogen Brackett series, and HeI emission line at $\lambda$2.06$\mu$m. The 2-D image shows Br$\gamma$ emission concentrated on the stellar source, while the H$_2$ is extended and coming from the surrounding nebulosity. Like Source C, temperatures $>$20\,000 K are required to excite the helium gas. The presence of both HeI similar to Source C suggests a late O-early B spectral type. Following the prescription for Source C, we adopt log\,$T_{\rm eff}$ of 4.5\,$^{+0.1}_{-0.2}$. The spectrum of ISO peak I shows strong PolyAromatic Hydrocarbon (PAH) 
typical of the mid-IR spectrum of a photo-dissociation region (PDR). We believe that
the H$_2$ emission is arising from the PDR (Whelan et al. 2013).

It should be noted that the LCO images show one source while the higher resolution ISAAC/VLT images 
resolve the source into two components separated by $\sim1\arcsec$. The spectrum shown in Figure 3 corresponds to the 
brightest IR component. The other IR component in the slit is a field star 
with a spectrum (not shown) typical of a carbon star. 

\subsection{Extinction}

We estimate the extinction of each source using the difference between the observed and intrinsic archival optical colours. We do not use the near-infrared colours as they are likely affected by strong infrared excesses (see Section 3.1). The intrinsic colours are calculated by convolving the appropriate filter curves with Castelli \& Kurucz (2004) stellar models at $Z$\,=\,0.004. We set the luminosity classification to a dwarf star, i.e. log\,$g$=4.0. Note that the luminosity classification (except if the star is a supergiant) does not affect the stellar colour within the error of the spectral typing. 

For Source C, we find $E(V-I)$\,=\,1. Based on the extinction coefficients of Schlegel et al. (1998), we calculate $A_V$=2.5\,$\pm$\,1, where the error arises from a combination of the error on the reddening due to the uncertainty in spectral type, and random photometric error. The nearest optical counterpart to Source E in Sabbi et al. (2007) is around 0.2$\arcsec$ from our nIR source, and does not seem to be physically related to it. Instead, we use the $J$-band magnitude to determine the extinction. We assume an intrinsic $J$ magnitude of a B0-B1 star \citep{Mamajek2012}, leading to $A_J$\,=\,0.45$\pm$0.4, where the error is purely due to the uncertainty in spectral type. This places the absolute extinction, $A_{V}$\,=\,1.7$\pm$1. Source I, has $E(V-I)$\,=\,0.5, leading to an $A_{V}$=1.2$\pm$1.

\subsection{Infrared SED}

The near-infrared and archival optical and infrared photometry was used to construct spectral energy distribution (SED) of each source (Fig. 4). The SED was then used to calculate the SED slope in the infrared, using the formula, 

\begin{equation} 
\alpha_{\rm{IR}}\,=\,\frac{d\,{\rm{log}}(\lambda F_{\lambda})}{d\,{\rm{log}}\lambda},      
\end{equation}
where $\lambda$ is between 3.6 and 24\,$\mu$m, and is used to diagnose the evolutionary stage of the disk-star system Lada (1987). We use the classification scheme of \cite{Greene94} to distinguish between systems with likely protostellar disks (Class I), optically thick circumstellar disks (Class II), or debris/transitional disk (Class III) objects. To calculate the SED slope, we fit the {\it Spitzer} [3.6], [4.5], [5.0], [8.0] and 24$\mu$m fluxes calculated from their unreddened magnitudes. We used the excess ratios from Sung et al. (2009) to derive the extinction coefficients in the {\it Spitzer} bands. The resultant $\alpha_{\rm{IR}}$ values of all stars fall $>$ 0.1, classifying them as Class I sources. 
The SEDs beyond $\lambda$ $>$ 24$\mu$m, are dominated by dust, which must have been evacuated in the immediate vicinity of the star, or be in the form of a circumstellar disc for the central star to be optically visible. 

In addition, we fit the infrared SED to the models of Robitaille et al. (2006). The model grid covers 20000 radiative transfer YSO models across a mass range of 0.1- 50$M_{\odot}$. From these fits, we may estimate approximately the stage of a YSO, it's mass and approximately luminosity with the associated fitting errors. We find that the best fit result ($\chi^2/datapoint<$4) for Source C corresponds  to a Stage I YSO, with a mass of 18$M_{\odot}$. Simon et al. (2007) classified the same source based purely on the mid-infrared photometry from {\it Spitzer} and found a fit consistent to either  an AGB star, or a YSO, favouring the later. For Source E, the best fit results in Stage I YSO, with a mass of 17\,$M_{\odot}$ closely matching that from Simon et al. (2007). The best-fit for Source I is extremely poor ($\chi^2/datapoint\sim$100), and we do not consider it a valid representative of the SED. It is shown for completeness. We stress that the SED fits are not intended to provide accurate parameters, but as a crude guide to the nature of each sources and that these models are limited in nature compared to the available free parameters (Offner et al. 2012).  

\begin{figure*}[t!]
\label{sed}
\centering
   \includegraphics[width=9cm]{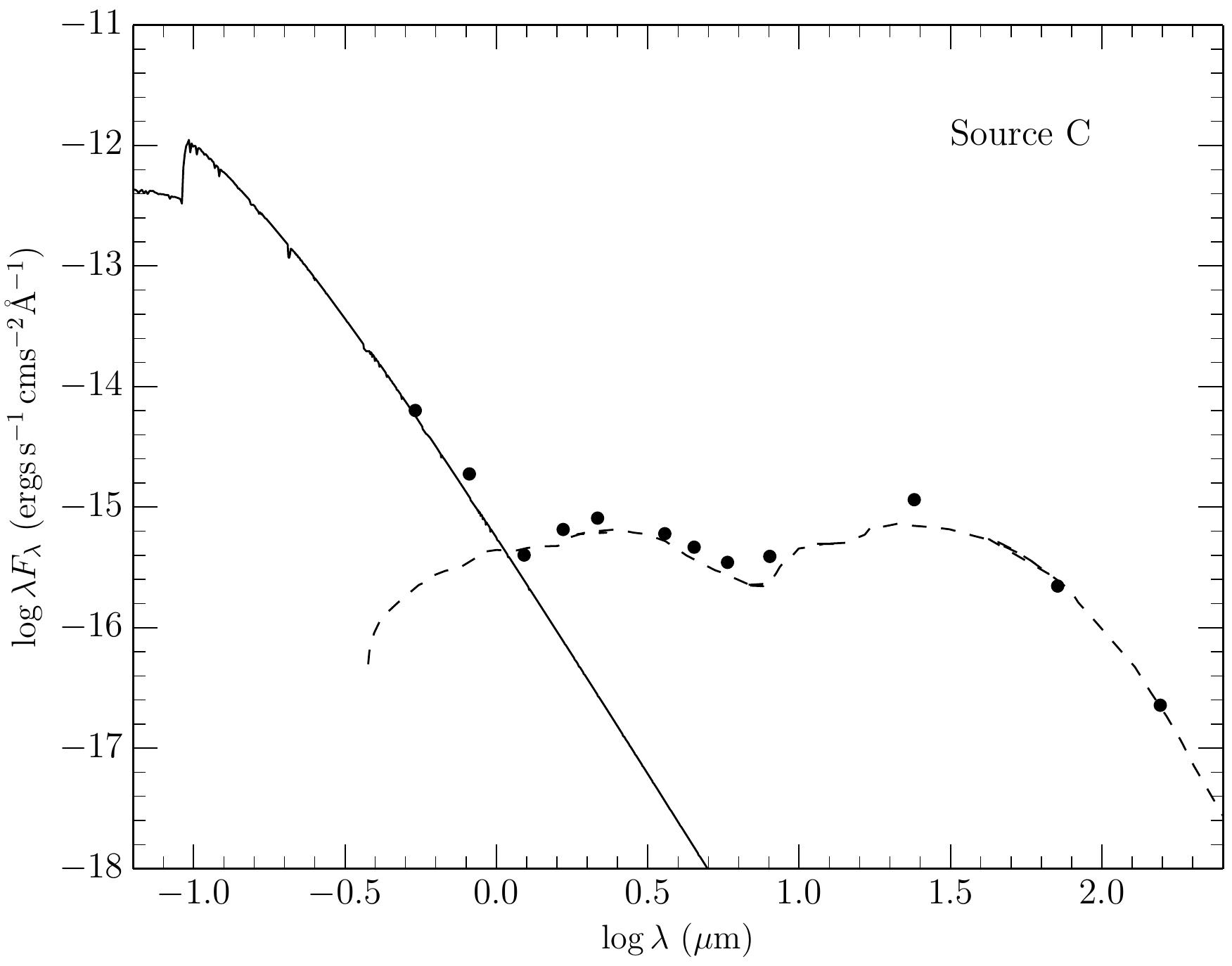}
   \includegraphics[width=9cm]{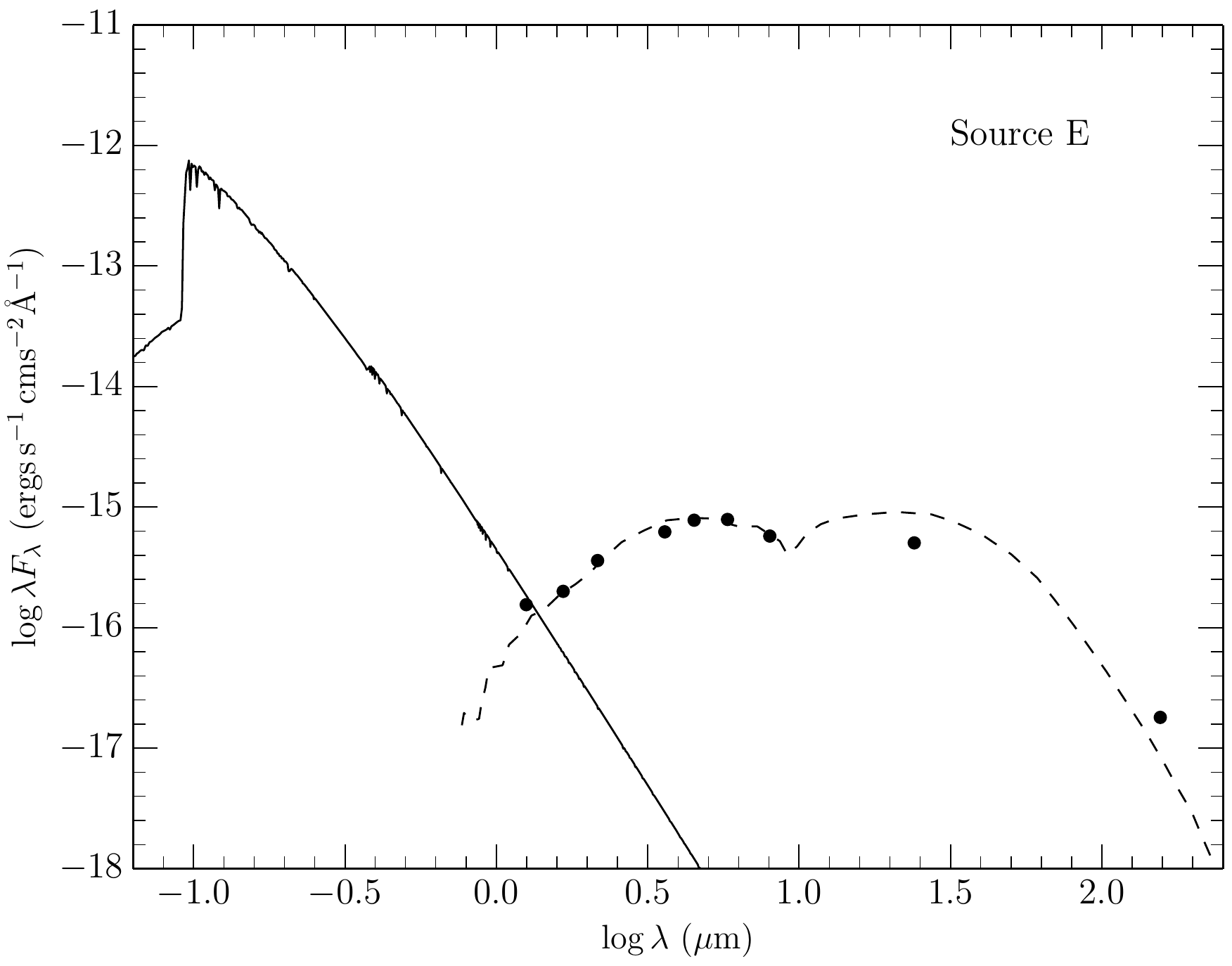}
   \includegraphics[width=9cm]{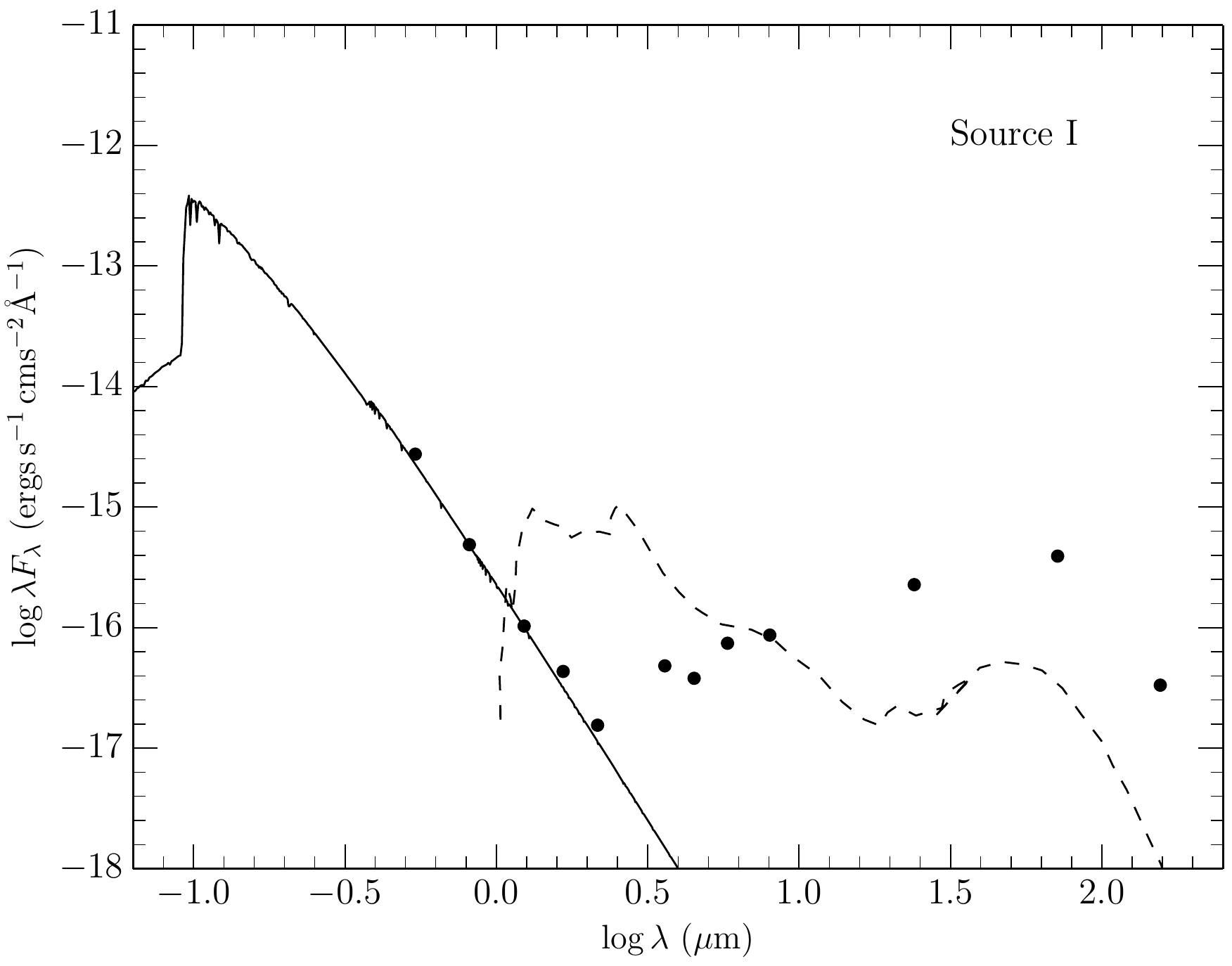}
\caption{Optical (0.55$\mu$m)-- mid infrared (160$\mu$m) SED of each of the sources. The overplotted stellar spectra (solid lines) are 1/5$Z_{\odot}$ Castelli \& Kurucz (2004) models at that nearest $T_{\rm eff}$ to that derived in Section 3.1. The best fitting YSO model from Robitaille et al. (2006) are also overplotted as dashed lines. The dots mark the photometry from Tables 2 \& 3, corrected for extinction values estimated in Section 5.2.}
\label{sed1}
\end{figure*}

\subsection{H$_2$ line ratios}

Molecular H$_2$ nIR emission in YSOs is caused either by (i) collisional excitation caused by outflows or shock from the YSOs (ii) UV radiation from nearby OB stars can form a photodissociation region (PDR). The flux-calibrated line ratios are an important diagnostic tools to differentiate between collisional and fluorescence in PDR regions (Black \& van Dishoeck 1987; Sternberg \& Dalgarno 1989; Draine
\& Bertoldi 1996, Burton et al. 1990, Burton et al. 1998). In the case of collisional excitation, the temperature of the gas is typically of order 2000--3000 K, resulting in only the lower levels of H$_2$ being populated. Whereas in the case of UV fluorescence, the H$_2$ molecule is lifted into high and low states by absorption of high-energy UV photons which subsequently decay.  Therefore, in the cases of collisional excitation, there are absence of transitions from high levels and a high ratio ($>$10) of fluxes in the 2.12$\mu$m 1-0 S(1) to the 2.24$\mu$m 2-1 S(1) lines, and a low ratio ($<$1) in the 2.22$\mu$m 1-0 S(0) to the 2.12 $\mu$m 1-0 S(1). On the contrary, in the case of UV fluorescence, there is a low ratio of fluxes ($<$2) in 2.12$\mu$m 1-0 S(1) to the 2.24$\mu$m 2-1 S(1), and a high ratio ($>$1) in 2.22$\mu$m 1-0 S(0) to the 2.12 $\mu$m 1-0 S(1).

The H$_2$ line fluxes for the three transitions given in Tables 4-6 shows that these are consistent with  models for dense PDRs. In source C, where we have the best signal-to-noise, we are dominated by an approximate pure fluorescence dense PDR regime, although we cannot discard the possibility of collisional excitation. In source E, the non-detection of the H$_2$ emission at 2.24 $\mu$m puts in a lower limit to the ratio of this transition. 
Given this fact, the H$_2$ emission would also originate in
dense PDRs with a greater degree of local extinction (ratio increases due to the increased intensity of H$_2$ 2.22$\mu$m) although we cannot rule out the presence of excitation by unresolved shocks in the same area. 

Excitation by shock is expected to happen very close to the mYSOs, at scales of 0.1-0.2 pc. While further away, the dominant mechanism should be the UV radiation from
the surrounding OB association (Israel \& Koorneef 1988).  Thus, it is very difficult to  resolve spatially the H$_2$ emission caused by shock or
 fluorescence at the distance of the SMC, where 1$\arcsec$=0.3 pc. We could most probably 
 be in the presence of a  combined contribution of emission mechanisms and therefore we can
 only consider the dominant one. In the case of source C, where the spectrum shows the best S/N ratio, the H$_2$ emission extends 2.5$\arcsec$ to each side of the source (0.8 pc) which would indicate the existence of a cocoon with fluorescence emission. The source C spectrum has HeI lines so it can provide the necessary 
UV photons so that an internal PDR  in the cocoon is formed. Nevertheless, we have to consider that 
source C is located near the core of NGC346 cluster, which  contains tens of massive stars, and
thus could  also contribute to the PDR H$_2$ emission.

\section{Discussion}

\subsection{Evolutionary status}

The evolutionary status of our stars is discerned using their positions in the Hertzsprung-Russell diagram (HRD), their infrared excess and SED, and their spectral signatures.

Firstly, we place them in the Hertzsprung-Russell diagram in Fig. 5. To calculate the luminosity, we applied bolometric corrections for the $J$ band calculated from the appropriate Castelli \& Kurucz (2004) model on the extinction and distance corrected $J$-band magnitude. The uncertainty on the luminosities includes the uncertainty in the extinction and spectral classification. We overplot PMS single star isochrones and stellar mass tracks at $Z$=0.004 (Bressan et al. 2012) up to the main-sequence phase on the HRD. Also shown are the location of known mYSO in the Galaxy, and the SMC.

The three sources lie above the main sequence, and between the 10,000 and 50,000 year isochrones, with estimated logarithm of ages, log\,$t$ of 4.25$\pm$1.0, 4.51$\pm$0.6, and 4.44$\pm$1.0 years for sources C, E, and I, respectively, from interpolating between the isochrones. These sources fall in the plane occupied by the brightest known SMC YSOs (Sewilo et al. 2013). Compared to the Galactic YSOs, for e.g. in W49, they are more luminous than their exact spectral type counterparts, although none are as massive as the $>$50 $M_{\odot}$ YSOs reported in that region (Wu et al. 2016). Interpolating from the mass tracks, we estimate that source C is the most massive, at around 26.0$\pm$5.0 $M_{\odot}$, and E and I at 15.9$\pm$3.0\,$M_{\odot}$ and 19.7$\pm$4.0\,$M_{\odot}$ respectively. Note the precise values of the masses and ages derived using both the isochrones and SED model fits are uncertain, and only provide an approximate estimate of both parameters. The final stellar parameters are tabulated in Table 7.

The 3$\mu$m-24$\mu$m SED slope of all three sources fall within Class I YSO classification, suggesting an infrared disc peaking at $\lambda$ $>$ 100$\mu$m, indicative of strong infrared excess. In addition, we fit the photometry having wavelengths greater than 3$\mu$m with the YSO models from Robitaille et al. (2006), with the results shown in Fig. 4. Source C \& E have good fits (with $\chi^2$/datapoint $<$4) for young massive YSOs, although Source I has a poor fit. These infrared characteristics are unique to YSOs, with the exception of B[e] stars that are a heterogeneous group of stars with infrared excesses similar to Class II YSOs, but B[e] stars have metal line emission (Liermann et al. 2014) in their spectra which are not seen in these spectra. All three  spectra show Br$\gamma$ emission close to the source, possibly from mass being accreted from a circumstellar disc, although the combination of the resolution and SNR do not allow for a detailed examination of the line profiles. In addition, all the sources show extended H$_2$ emission, arising mostly due to fluorescence, but there could be also a possible contribution from shocked gas from outflows. All these characteristics point unambiguously towards the classification as mYSOs. 

\begin{figure}
\label{sed}
\centering
\includegraphics[width=8cm, angle=0]{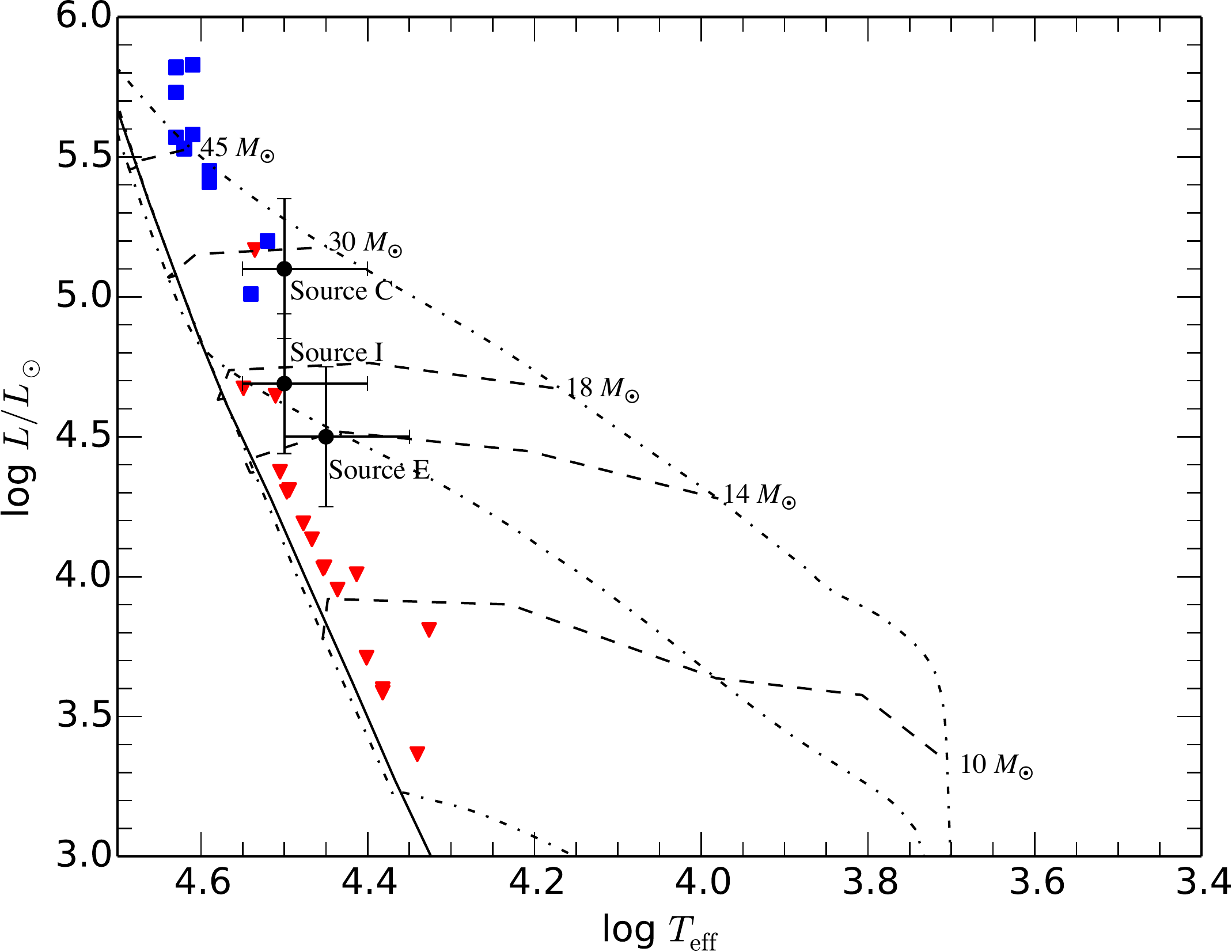}
\caption{The position of our mYSO candidates (circles) are shown along with known SMC YSOs (Sewilo et al. 2013) as red inverted triangles and galactic YSOs (Wu et al. 2016) as blue squares. The dashed dotted lines are the PMS isochrones of Bressan et al. (2012) at 1/5$Z_{\odot}$ for 0.01, 0.03 and 0.3 Myr right to left respectively, and the dashed line the stellar tracks for 10, 14, 18, 30 and 45 $M_{\odot}$ from bottom up. The solid line is the main-sequence locus. }
\label{sed1}
\end{figure}

\subsection{Comparison with literature classifications}

There are two major YSO infrared studies conducted using data from the {\em Spitzer} telescope which include our three sources- Sewilo et al. (2013) and Simon et al. (2007).

Sewilo et al. (2013) classified {\it Spitzer} YSO candidates in the SMC using 3.6--24 $\mu$m photometry. Their classification was based on the position of each candidate in various infrared CMD and CCD, which allowed them to differentiate them from contaminants. Based on their positions in multiple diagrams, each candidate was given a score indicating the likelihood of being a YSO. In addition, they also fit SED models to their photometry, however none of the SED fits in Sewilo et al. (2013) meeting their $\chi^2$ requirement include our sources. Instead, Source C (Y535) and E (Y540) are both classified as possible YSO candidates based on their CMD scores, while source I (Y552) is classified as YSO, and is flagged as likely having PolyAromatic Hydrocarbon (PAH) emission based on its SED fit, which contributes to its poor fit score. The  emission is confirmed in the study done by Whelan et al. (2013) using Spitzer mIR spectra.

Simon et al. (2007) have catalogued a list of YSO's in N66 based on the
\textit{Spitzer} photometry and modelling adjusted to
the obtained spectral energy distribution (SED).
Source E is the brightest source of their catalogue (their source 25),
with a luminosity of $30\,300$\,L$_\odot$ and a stellar mass of
$16.6$\,M$_\odot$.  
The source is classified as Stage I type, following the classification scheme
defined by Robitaille et al. (2006). 
Source C is mentioned as the fourth most luminous object at $8\mu$m,
i.e. SSTS3MC\,$14.7725-72.1766$ but they do not classify it as a YSO as they 
could fit its SED either with a YSO or an AGB SED. They favour the possibility 
of it being a YSO with a luminosity of $3.3\,10^4$\,L$_\odot$ and mass of
$14.7$\,M$_\odot$, mainly based on the fact that the source is located in
the center of NGC 346 and surrounded by many young massive stars. The luminosities of both sources from Simon et al. (2007) fall within the error ranges of the luminosities derived in our work, however the mass of only source E matches our derived mass within error estimates, whereas Source C, we find a masses $\sim$ 10$M_{\odot}$ higher than that estimated from SED fitting in Simon et al. (2007). From the spectrum alone based on the HeI line, we suggest that C should be much more massive than derived in Simon et al. (2007). Source I is not in their catalogue.  

\subsection{Formation scenario}

The molecular gas observations by  Rubio et al. (2000) and the mid-IR ISO
observations by  Contursi et al. (2000) suggested that the dense clumps of 
neutral hydrogen gas and dust in the N66 HII region could be the sites of recent star 
formation. 
The IR spectra obtained towards sources C, E and I in N66 plus their nIR
colours confirm that these sources are embedded mYSOs with characteristics of the Class I type, thereby confirming that these dense clumps harbour YSOs, and are the sites of recent star formation. 
They are located in regions of high density of neutral molecular gas as seen
in the CO(2-1) emission line and are associated to H$_2$ knots as seen in the
H$_2$ images as well as spatially resolved emission in the long-slit nIR
spectra. 

From the derived properties of the sources, we find that source C is the most massive with a mass $>$25\,$M_{\odot}$, whereas sources E and I are almost at similar ages, with masses around $\sim$ 15--20 $M_{\odot}$. This suggests a different scenario to the sequential star formation scenario of Contursi et al. (2000) and Rubio et al. (2000), where the authors suggest that star formation has begun at NGC346 and continues outward along the bar. Instead, the youngest and most massive YSO is found in the central cluster, while comparatively older and less massive YSOs are found along the bar.
 

Although, the low numbers do not enable any statistical study for age spreads or a formation scenario, we can argue that star formation at least in the central cluster NGC 346 seems to have  been ongoing. Source C, which has moderate extinction of $A_V$ $\sim$2.5, is not located behind the molecular clump (which would suggest an extinction $>$8), so it should lie at the cluster, or between cluster and our line of sight. Therefore it is most likely part of the cluster, which has literature age of $\sim$3 Myr (Sabbi et al. 2007) based on isochrone fitting. De Marchi et al. (2013) find a much younger PMS population identified based on their H$\alpha$ excess emission of around 1 Myr, which is more in line with the age we estimate for Source C.  The presence of a mYSO at the very heart of the cluster, suggests that it must have formed there indicating that star formation has either been on-going continuously, or has undergone another very recent burst since the formation of the cluster four million years ago. 
It is interesting to note that the H$\alpha$ HST image shows a source associated to the nIR source C. Infact, all our information suggests that source C is a complex source, and is a likely a mYSO enshrouded in a compact HII region excited by O stars that ionize  the hydrogen gas around them. The detected H$_2$ lines could be produced on the interface of the PDR where the neutral gas is being exited by these young stars. Molecular gas is present in the region although we do not presently  have a  high resolution CO map towards NGC346 to show the molecular cloud structure. This structure  most probably would consist on dense  and small CO cores as those  seen with ALMA arc second resolution CO images near 30 Doradus (Indebetouw et al. 2013) or in WLM (Rubio et al. 2015). Further, arcsecond resolution molecular maps are necessary to confirm this hypothesis.

\begin{table}
\begin{center}
\caption{Identified lines of source C\label{tbl-1}}
\begin{tabular}{lccc}
\hline\hline
line & $\lambda$ & Flux & EW\\ 
     & $(\mu$m)	  &  (erg~$cm^{-2} s^{-1}$) & (\AA)\\
\hline
\hline
  Br19 & 15264.7 & 2.3$\times10^{-15}$ & -3.8\\
  Br18 & 15346.0 & 1.5$\times10^{-15}$ & -2.4\\
  Br17 & 15443.1 & 2.1$\times10^{-15}$ & -3.5\\
  Br16 & 15560.7 & 3.4$\times10^{-15}$ & -3.4\\
  Br15 & 15705.0 & 3.4$\times10^{-15}$ & -5.8\\
  Br14 & 15884.9 & 4.9$\times10^{-15}$ & -8.3\\
  Br13 & 16113.7 & 6.7$\times10^{-15}$ & -11.5\\
  Br12 & 16411.7 & 7.9$\times10^{-15}$ & -13.6\\
  Fe$[II]$ & 16434.0 & 2.6$\times10^{-15}$ & -4.5\\
  Br11 & 16811.1 & 1.0$\times10^{-14}$ & -17.0\\
  Fe$[II]$ & 16877.8 & 3.0$\times10^{-15}$ & -5.1\\
  Br10 & 17366.7 & 1.34$\times10^{-14}$ & -22.5\\
  H$_2$ (1-0)\,S(7) & 17480.0 & 3.4$\times10^{-15}$ & -5.6\\
  Br9 & 18179.1 & 1.27$\times10^{-14}$ & -20.0\\
  Pa4 & 18756.1 & 8.02$\times10^{-14}$ & -134.0\\
  Br8 & 19450.9 & 1.79$\times10^{-14}$ & -27.7\\
  H$_2$ (1-0)\,S(3) & 19576.0 & 4.2$\times10^{-15}$ & -6.4\\
  CaI & 19782.0 & 7.0$\times10^{-16}$ & -1.1\\
  Fe$[II]$ blend & 20041.1 & 5.1$\times10^{-15}$ & -7.7\\
  H$_2$ (1-0)\,S(2) & 20338.0 & 1.8$\times10^{-15}$ & -2.7\\
  HeI & 20587.0 & 1.3$\times10^{-15}$ & -1.9\\
  H$_2$ (2-1)\,S(3) & 20730.0 & 1.2$\times10^{-15}$ & -1.7\\
  H$_2$ (1-0)\,S(1) & 21218.3 & 8.4$\times10^{-15}$ & -11.6\\
  Br7 (Br$\gamma$) & 21661.3 & 2.29$\times10^{-13}$ & -32.3\\
  NaI & 22062.4 & 9.0$\times10^{-16}$ & -1.2\\
  H$_2$ (1-0)\,S(0) & 22235.0 & 3.3$\times10^{-15}$ & -4.7\\
  H$_2$ (2-1)\,S(1) & 22477.0 & 5.1$\times10^{-15}$ & -7.0\\
  H$_2$ (1-0)\,Q(1) & 24066.0 & 1.13$\times10^{-14}$ & -14.6\\
  H$_2$ (1-0)\,Q(2) & 24134.0 & 1.8$\times10^{-15}$ & -2.4\\
  H$_2$ (1-0)\,Q(3) & 24237.0 & 2.5$\times10^{-15}$ & -3.2\\
\hline\end{tabular}
\end{center}
\end{table}

\begin{table}
\begin{center}
\caption{Identified lines of source E\label{tbl-1}}
\begin{tabular}{lccc}
\hline\hline
line & $\lambda$ & Flux & EW\\ 
     & $(\mu$m)	  &  (erg~$cm^{-2} s^{-1}$) & (\AA)\\
\hline
\hline
  Br13 & 16113.7 & 3.7$\times10^{-15}$ & -14.8\\
  Fe$[II]$ & 16434.0 & 2.8$\times10^{-15}$ & -11.6\\
  Br10 & 17366.7 & 5.6$\times10^{-15}$ & -21.8\\
  Br8 & 19450.9 & 5.8$\times10^{-15}$ & -17.7\\
  H$_2$ (1-0)\,S(3) & 19576.0 & 3.9$\times10^{-15}$ & -11.1\\
  Fe$[II]$ blend & 20041.1 & 2.3$\times10^{-15}$ & -6.4\\
  H$_2$ (1-0)\,S(2) & 20338.0 & 2.4$\times10^{-15}$ & -6.3\\
  HeI & 20587.0 & 1.3$\times10^{-15}$ & -1.9\\
  H$_2$ (2-1)\,S(3) & 20730.0 & 3.1$\times10^{-15}$ & -7.9\\
  H$_2$ (1-0)\,S(1) & 21218.3 & 6.8$\times10^{-15}$ & -17.2\\
  Br7 (Br$\gamma$) & 21661.3 & 5.4$\times10^{-15}$ & -12.7\\
  H$_2$ (1-0)\,S(0) & 22235.0 & 1.9$\times10^{-15}$ & -4.3\\
  H$_2$ (1-0)\,Q(1) & 24066.0 & 9.9$\times10^{-15}$ & -16.4\\
  H$_2$ (1-0)\,Q(2) & 24134.0 & 2.5$\times10^{-15}$ & -4.2\\
  H$_2$ (1-0)\,Q(3) & 24237.0 & 3.5$\times10^{-15}$ & -5.7\\
  H$_2$ (1-0)\,Q(4) & 24375.0 & 2.6$\times10^{-15}$ & -4.3\\
\hline\end{tabular}
\end{center}
\end{table}

\begin{table}
\begin{center}
\caption{Identified lines of source I\label{tbl-1}}
\begin{tabular}{lccc}
\hline\hline
line & $\lambda$ & Flux & EW\\ 
     & $(\mu$m)	  &  (erg~$cm^{-2} s^{-1}$) & (\AA)\\
\hline
\hline
  KI & 15163.1 & 1.3$\times10^{-15}$ & -3.5\\
  Br19 & 15264.7 & 1.3$\times10^{-15}$ & -3.6\\
  FeI & 15299.0 & 0.7$\times10^{-15}$ & -2.1\\
  Br18 & 15346.0 & 0.7$\times10^{-15}$ & -2.1\\
  Br17 & 15443.1 & 0.7$\times10^{-15}$ & -1.9\\
  Br16 & 15560.7 & 1.5$\times10^{-15}$ & -4.0\\
  FeI & 15697.0 & 1.2$\times10^{-15}$ & -3.3\\
  Br15 & 15705.0 & 0.9$\times10^{-15}$ & -2.5\\
  MgI & 15753.0 & 1.2$\times10^{-15}$ & -3.6\\
  MgI+Br14 & 15885.0 & 0.9$\times10^{-15}$ & -2.7\\
  SiI+FeI & 15964.0 & 1.5$\times10^{-15}$ & -4.2\\
  SiI+Br13 & 16099.0 & 4.4$\times10^{-15}$ & -12.4\\
  SiI+FeI & 16386.0 & 2.4$\times10^{-15}$ & -7.1\\
  FeII & 16440.0 & 3.7$\times10^{-15}$ & -10.8\\
  Br11 & 16811.2 & 4.6$\times10^{-15}$ & -13.0\\
  CI & 16895.0 & 0.9$\times10^{-15}$ & -2.6\\
  MgI & 17113.3 & 0.9$\times10^{-15}$ & -2.6\\
  Br10+NaI & 17366.7 & 4.4$\times10^{-15}$ & -11.7\\
  H$_2$ (1-0)\,S(7) & 17480.0 & 3.4$\times10^{-15}$ & -8.9\\
  H$_2$ (1-0)\,S(2) & 20338.0 & 2.7$\times10^{-15}$ & -6.5\\
  HeI & 20587.0 & 10.1$\times10^{-15}$ & -24.5\\
  H$_2$ (2-1)\,S(3) & 20730.0 & 5.0$\times10^{-15}$ & -12.7\\
  H$_2$ (1-0)\,S(1) & 21218.3 & 12.0$\times10^{-15}$ & -32.2\\
  H$_2$ (2-1)\,S(2) & 21542.0 & 1.4$\times10^{-15}$ & -3.9\\
  Br7 (Br$\gamma$) & 21661.3 & 24.0$\times10^{-15}$ & -65.1\\
  HeI & 21840.0 & 1.8$\times10^{-15}$ & -5.0\\
  NaI & 22062.4 & 2.4$\times10^{-15}$ & -6.6\\
  NaI & 22089.7 & 3.4$\times10^{-15}$ & -9.4\\
  H$_2$ (1-0)\,S(0) & 22235.0 & 3.0$\times10^{-15}$ & -8.3\\
  TiI & 22280.1 & 3.6$\times10^{-15}$ & -9.9\\
  H$_2$ (2-1)\,S(1) & 22477.0 & 2.2$\times10^{-15}$ & -6.1\\
  CaI & 22614.1 & 5.2$\times10^{-15}$ & -14.7\\
  CaI & 22657.3 & 46.0$\times10^{-15}$ & -17.5\\
  FeII & 22727.3 & 4.0$\times10^{-15}$ & -12.8\\
  MgI & 22814.1 & 2.3$\times10^{-15}$ & -7.8\\
\hline\end{tabular}
\end{center}
\end{table}




\begin{table}
\begin{center}
\caption{Stellar parameters of the mYSOs}
\begin{tabular}{lcccc}
\hline\hline
Star & $T_{\rm eff}$ & log\,$L_{\ast}$ & Mass & Log\,Age \\
     & (K)  & ($L_{\odot}$) & ($M_{\odot}$) & (yr) \\
\hline
Source C &  4.5\,$^{+0.1}_{-0.2}$    &   5.1$\pm$0.25     &  26$\pm$5    &   4.25$\pm$1.0 \\
Source E &  4.45\,$^{+0.1}_{-0.2}$  &   4.5$\pm$0.25      &  15.9$\pm$3.0	&   4.51$\pm$0.6 \\
Source I &  4.5\,$^{+0.1}_{-0.2}$  &    4.69$\pm$0.25     &  19.7$\pm$4.0   & 4.44$\pm$1.0 \\
\hline\end{tabular}
\end{center}
\end{table}

\section{Conclusions}

We present $HK$ spectra of three bright sources selected using nIR $JHK$ photometry in the N66/NGC346 region of the SMC. Based on a combination of the spectral features seen in the near-infrared, photometry and SED we suggest that Source C and I are of early spectral types, between late O to early B.  Source E is more likely a early B star. This spectral classification provides a crude estimate of the effective temperature and extinction of the sources which allows us to place them in the HRD. Based on their positions in the HRD we derive that they have ages between 10000 and 50000 years. The evolutionary mass of Source C would be $\sim$25\,$M_{\odot}$, while Source E and I, is between 15--20\,$M_{\odot}$. The SEDs of each source shows excess longwards of $\lambda$$>$1$\mu$m. Both, their spectral energy distribution slope and results from fitting against the YSO models of Robitaille et al. (2006) suggest that these three sources are likely massive young stellar objects. The nature of source C is more complex compared to source E and I, given its high nIR luminosity and position near the centre of NGC\,346, and further studies are necessary to understand this source. Combining the complete information at our disposal, we suggest that sources observed here are massive young stellar objects in N\,66. The nIR spectroscopy combined with nIR-mid IR photometry, confirms that a new generation of stars is  being formed in the dense clumps in the neutral molecular gas of N66. The interaction of the massive stars in NGC\,346,  
with the surrounding molecular gas may have created the conditions for a new star formation event.



\begin{acknowledgements}

We would like to thank the anonymous referee for his/her valuable comments and careful reading of the manuscript. M.R. wishes to acknowledge support from CONICYT (CHILE) through FONDECYT grant No1140839 and partial support through
 project BASAL PFB-06.
 R.B. wishes to acknowledge support from  
FONDECYT (CHILE) grant No 1050052, and DIULS CD 08102, and V.M.K acknowledges support from the FONDECYT Postdoc Fellowship grant No. 3116017. 

\end{acknowledgements}

\end{document}